\documentclass[11pt]{article}
\usepackage{epsfig,dsfont}
\usepackage{geometry,wasysym}
\begin{document}  
\sffamily

\vspace*{1mm}

\begin{center}

{\LARGE
Kramers-Wannier duality and worldline representation\\
\vskip2mm
 for the SU(2) principal chiral model}
\vskip15mm
Christof Gattringer, Daniel G\"oschl, Carlotta Marchis
\vskip5mm
Universit\"at Graz, Institut f\"ur Physik, Universit\"atsplatz 5, 8010 Graz, Austria 
\end{center}
\vskip12mm

\begin{abstract}
In this letter we explore different representations of the SU(2) principal chiral model on the lattice. We couple chemical potentials to 
two of the conserved charges to induce finite density. This leads to a complex action such that the conventional field 
representation cannot be used for a Monte Carlo simulation. Using the recently developed Abelian color flux approach we 
derive a new worldline representation where the partition sum has only real and positive weights, such that a Monte Carlo simulation 
is possible. In a second step we transform the model to new dual variables in the Kramers-Wannier (KW) sense, such that the 
constraints are automatically fulfilled, and we obtain a second representation free of the complex action problem. We implement 
exploratory Monte Carlo simulations for both, the worldline, as well as the KW-dual form, for cross-checking the two dualizations and a
first assessment of their potential for dual simulations. 
\end{abstract}

\vskip14mm

\section{Introductory remarks}
For many (lattice) field theories the introduction of a chemical potential leads to a complex action $S$ such that the Boltzmann 
factor exp$(-S)$ of the path integral cannot be used as a weight in a Monte Carlo simulation. Considerable progress towards 
overcoming this so-called complex action problem was made by exactly rewriting models to a worldline representation (worldsheets
for gauge degrees of freedom), where the partition sum has only real and positive contributions, such that a Monte Carlo 
simulation is possible in terms of the new degrees of freedom (for an overview of these developments see, e.g., 
the reviews at the annual lattice conferences \cite{reviews}). 

So far simulations were mostly done directly in the worldline representation using algorithms that deal with the constraints, often 
based on the Prokofev-Svistunov worm algorithm \cite{worm}. However, in some cases it is possible to get rid of the constraints by a 
second transformation to dual variables in the Kramers-Wannier (KW) sense (see \cite{savit} for a review). One represents the worldlines
with new dual variables that automatically obey the constraints. These new fully dual representations also interchange the weak and
strong coupling regions of the original formulation in terms of field variables and might turn out to be a better choice for simulations 
in some parameter regions. Furthermore they might open the way to yet another possible simulation strategy, namely using 
Swendsen-Wang or Wolff cluster algorithms \cite{cluster}. 

In this letter we present a new worldline representation for the SU(2) principal chiral model where we couple chemical potentials to 
two of the conserved charges. This allows one to study finite density of these charges, but also provides insight how 
symmetries of the conventional field representation manifest themselves in the worldline- and KW-dual representations. Our 
worldline representation uses the recently developed Abelian color flux approach \cite{abelian_cf}, where the worldlines propagate in both, 
the space-time lattice, as well as color space (or more generally the space of indices of the symmetry group of a model). In a 
second step we derive the full KW-dualization by introducing variables for local plaquette flux and variables for disorder flux that winds
around the spatial and temporal boundaries. These dual variables are free of constraints. In an exploratory implementation we perform a
Monte Carlo simulation of the worldline and the KW-dual form. This mainly serves to cross-check the transformations presented here,
but also allows for a first assessment of Monte Carlo simulations in the different forms. 

\section{Continuum model and conventional lattice form}

In the continuum the SU(2) principal chiral model in $d$ dimensions and with coupling $J$ is described by the Euclidean continuum 
action 
\begin{equation}
S \;  =  \; \frac{J}{2} \int \! \textnormal{d}^dx \; \textnormal{Tr}\left[ (\partial_\nu U(x))^\dagger (\partial_\nu U(x)) \right] \; , 
\end{equation}
where the dynamical degrees of freedom $U(x)$ are elements of SU(2) and the Euclidean space-time indices $\nu$ are summed  
over $\nu = 1,2\, ... \, d$. The action is invariant under the global symmetries of multiplying $U(x)$ with arbitrary SU(2) matrices from 
left and/or right. For each of these the corresponding Noether charges can be determined and coupled with chemical
potentials. Here we study two of these symmetries ($\sigma_3$ denotes the third Pauli matrix),
\begin{equation}
U(x) \; \rightarrow  \;	e^{\, i \frac{\alpha_1}{2} \sigma_3} \, U(x) \, e^{\, i \frac{\alpha_1}{2} \sigma_3}
\quad , \quad
U(x) \; \rightarrow  \;	e^{\, i \frac{\alpha_2}{2} \sigma_3} \, U(x) \, e^{\, -i \frac{\alpha_2}{2} \sigma_3} \; ,
\end{equation}
where $\alpha_1 \in \mathds{R}$ and $\alpha_2 \in \mathds{R}$ 
are independent parameters.  The corresponding Noether charges are 
\begin{eqnarray}
Q_1 & = & \frac{i J}{4} \int \! \textnormal{d}^{d-1}x \;  \textnormal{Tr} \Big[ \bigl(\partial_d \, U^\dagger(x)\bigr) 
[ \sigma_3 U(x) + U(x) \sigma_3] - U^\dagger(x) [ \sigma_3 (\partial_d \, U(x)) + (\partial_d \, U(x)) \sigma_3 ] \Big] \, ,
\nonumber \\
Q_2 & = & \frac{i J}{4} \int \! \textnormal{d}^{d-1}x \;  \textnormal{Tr} \Big[ \bigl(\partial_d \, U^\dagger(x)\bigr) 
[ \sigma_3 U(x) - U(x) \sigma_3] - U^\dagger(x) [ \sigma_3 (\partial_d \, U(x)) - (\partial_d \, U(x)) \sigma_3 ] \Big] \, ,
\label{noether}
\end{eqnarray}
where the integration runs only over the $d-1$ dimensional space.

The lattice version of the model with chemical potentials $\mu_1$ and $\mu_2$ coupled to the two charges is defined by
the action
\begin{equation}
S  =  - \frac{J}{2}  \sum_{x, \nu} \! \left( \!
\textnormal{Tr} \! \left[ e^{\, \delta_{\nu,d} \sigma_3 \frac{\mu_1 + \mu_2}{2}} \, U_x \, 
e^{\, \delta_{\nu,d} \sigma_3 \frac{\mu_1 - \mu_2}{2}} \, U_{x+\hat{\nu}}^\dagger \right] +  
\textnormal{Tr} \! \left[ e^{- \delta_{\nu,d} \sigma_3 \frac{\mu_1 - \mu_2}{2}} \, U_x^\dagger \,
e^{- \delta_{\nu,d} \sigma_3 \frac{\mu_1 + \mu_2}{2}} \, U_{x+\hat{\nu}} \right] \right) \! .
\label{latticeaction}
\end{equation}
Here the SU(2) matrices $U_x$ sit on the sites $x$ of a $d$-dimensional $N^{d-1} \times N_t$ lattice with periodic boundary conditions
and lattice constant set to $a = 1$. In the lattice discretization
the chemical potentials $\mu_1$ and $\mu_2$ couple in the familiar form, giving different weights to temporal ($\nu = d$) forward
and backward nearest neighbor terms, i.e., they give different weight to matter and anti-matter of the charge they couple to. 
It is easy to see (compare below), that non-zero values of the $\mu_\lambda$ give rise to a complex action $S$ and that in the 
conventional representation the model suffers from the complex action problem.

The partition function of the model is given by $Z = \int D[U] e^{-S}$, where $\int DU = \prod_x \int_{SU(2)} \textnormal{d} U_x$ 
denotes the product over all lattice sites $x$ of the SU(2) Haar-measure integrals $\int_{SU(2)} \textnormal{d} U_x$. 
In the following we will use the explicit representation 
($\theta_x \in [0,\pi/2], \, \alpha_x \in [-\pi, \pi], \, \beta_x \in [-\pi, \pi] $)
\begin{equation}
U_x \; = \; \left[ \begin{array}{cc} 
\cos \theta_x e^{\, i \alpha_x} & \sin \theta_x e^{\, i \beta_x} \\
-\sin \theta_x e^{\, - i \beta_x} & \cos \theta_x e^{\, - i \alpha_x} 
\end{array} \right] \quad \mbox{with} \quad \textnormal{d}U_x \; = \; 2 \sin \theta_x \, \cos \theta_x \, \textnormal{d} \theta_x \, 
\frac{\textnormal{d} \alpha_x}{2\pi} \frac{\textnormal{d} \beta_x}{2\pi} \; .
\label{parameterization} 
\end{equation}

\section{Worldline representation of the model}
The first step in our dualization program is the derivation of the worldline representation using the abelian color flux (ACF) approach 
\cite{abelian_cf}. For the ACF we write the traces and matrix products in the lattice action (\ref{latticeaction}) as explicit sums over 
the indices $a,b$ of the matrices $U_x^{ab}$. We find the following expression for $Z$:
\begin{eqnarray}
Z &\!\! = \!\!\!& \int \!\! D[U] \exp \! \left( \! J  \sum_{x,\nu} \! \bigg[ 
e^{\mu_1 \delta_{\nu,d}} \ U_x^{11}  {U_{x+\hat{\nu}}^{11}}^{\!\!\! \star} + 
e^{\mu_2 \delta_{\nu,d}} \ U_x^{12}  {U_{x+\hat{\nu}}^{12}}^{\!\!\! \star} +
e^{-\mu_2 \delta_{\nu,d}} \ U_x^{21}  {U_{x+\hat{\nu}}^{21}}^{\!\!\! \star} + 
e^{-\mu_1 \delta_{\nu,d}} \ U_x^{22}  {U_{x+\hat{\nu}}^{22}}^{\!\!\! \star} 
\bigg] \! \right) \nonumber \\
&\!\! = \!\!\!& \int \!\! D[U] 
\exp \! \left( \! J  \sum_{x,\nu} \sum_{a,b = 1}^2 M_\nu^{ab} U_x^{ab}  {U_{x+\hat{\nu}}^{ab}}^{\!\!\! \star} \right)
\; = \; \int \!\! D[U] \prod_{x,\nu} \prod_{a,b = 1}^2 \, e^{\, J M_\nu^{ab} U_x^{ab}  {U_{x+\hat{\nu}}^{ab}}^{\!\!\! \star} }\; ,
\label{trafo1}
\end{eqnarray} 
where in the second step we have introduced the matrices $M_\nu$ with entries 
\begin{equation}
M_ \nu^{11} = e^{\, \mu_1 \delta_{\nu,d} } \quad , \quad M_ \nu^{22} = e^{\, -\mu_1 \delta_{\nu,d} } \quad , \quad
M_ \nu^{12} = e^{\, \mu_2 \delta_{\nu,d} } \quad , \quad M_ \nu^{21} = e^{\, -\mu_2 \delta_{\nu,d}} \; ,
\end{equation}
and in the third step of (\ref{trafo1}) the exponential of sums was written as a product over exponentials of the individual terms. 
The next step for determining the worldline representation is to expand each of the exponentials  
$e^{\, J M_\nu^{ab} U_x^{ab}  {U_{x+\hat{\nu}}^{ab}}^{\!\!\! \star}}$ in a power series and to subsequently integrate out the 
$U_x^{ab}$. Introducing summation variables $j_{x,\nu}^{ab}$ for the power series for each of the exponentials we find
\begin{eqnarray}
Z &\!\! = \!\!\!& \int \!\! D[U] \prod_{x,\nu} \prod_{a,b} \sum_{j_{x,\nu}^{ab} = 0}^\infty 
\frac{ ( J M_\nu^{ab} )^{j_{x,\nu}^{ab}}}{j_{x,\nu}^{ab} \, !} \, \big( U_x^{ab}  {U_{x+\hat{\nu}}^{ab}}^{\!\!\! \star} \big)^{j_{x,\nu}^{ab}} 
\, = \; \sum_{\{ j \}} W_{J,\mu}[j] \int \!\! D[U] \prod_{x,\nu} \prod_{a,b} 
\big( U_x^{ab} \big)^{j_{x,\nu}^{ab}}   \big({U_{x}^{ab}}^{\star} \big)^{j_{x-\hat{\nu},\nu}^{ab}}
\nonumber \\ 
&\!\! = \!\!\!& \sum_{\{ j \}} W_{J,\mu}[j] \, \prod_x 2 \!\! \int_0^{\frac{\pi}{2}} \!\! \textnormal{d}\theta 
(\cos \theta)^{1 + \sum_\nu\big[ j_{x,\nu}^{11} + j_{x,\nu}^{22} + j_{x-\hat{\nu},\nu}^{11} + j_{x-\hat{\nu},\nu}^{22}\big]} \,
(\sin \theta)^{1 + \sum_\nu\big[ j_{x,\nu}^{12} + j_{x,\nu}^{21} + j_{x-\hat{\nu},\nu}^{12} + j_{x-\hat{\nu},\nu}^{21}\big]}
\nonumber \\
&& \hspace{5mm} \times \int_{-\pi}^{\pi} \frac{\textnormal{d}\alpha}{2\pi} \,
e^{\, i \alpha \sum_\nu\big[ \big( j_{x,\nu}^{11} - j_{x,\nu}^{22}\big) - \big(j_{x-\hat{\nu},\nu}^{11} - j_{x-\hat{\nu},\nu}^{22}\big)\big]}
 \int_{-\pi}^{\pi} \frac{\textnormal{d}\beta}{2\pi} \,
e^{\, i \beta \sum_\nu\big[ \big( j_{x,\nu}^{12} - j_{x,\nu}^{21}\big) - \big(j_{x-\hat{\nu},\nu}^{12} - j_{x-\hat{\nu},\nu}^{21}\big)\big]} \; .
\label{trafo2}
\end{eqnarray}
In the second step we have reordered the factors $U_x^{ab}$ and introduced the sum $\sum_{\{ j \}}$ 
over all configurations of the summation variables $j_{x,v}^{ab}$, as well as the weight factors
$W_{J,\mu}[j]$:
\begin{equation}
\sum_{\{ j \}} \, \equiv \, \prod_{x,\nu} \prod_{a,b} \sum_{j_{x,\nu}^{ab} = 0}^\infty \, , \;
W_{J,\mu}[j] \, \equiv \, \prod_{x,\nu} \prod_{a,b} \frac{ ( J M_\nu^{ab} )^{j_{x,\nu}^{ab}}}{j_{x,\nu}^{ab} \, !} \, = \, 
e^{\, \mu_1 \! \sum_x \! \big[ j_{x,d}^{11} - j_{x,d}^{22} \big]}
e^{\, \mu_2 \! \sum_x \! \big[ j_{x,d}^{12} - j_{x,d}^{21} \big]}
\prod_{x,\nu} \prod_{a,b} \frac{J^{j_{x,\nu}^{ab}}}{j_{x,\nu}^{ab} \, !} \; .
\end{equation}
In the third step of (\ref{trafo2}) we have inserted the path integral measure $D[U]$ and the explicit parameterization
(\ref{parameterization}) for the $U_x$ and the Haar integration measure (we have dropped the indices $x$ of $\theta_x, \, 
\alpha_x$ and $\beta_x$ for better readability). The integrals in (\ref{trafo2}) can be solved in closed form: 
The integrals over $\alpha$ and $\beta$ give rise to Kronecker deltas for the integer valued combinations of the $j_{x,\nu}^{ab}$ in 
the respective exponents, i.e., these intergrals generate constraints for these combinations at each site $x$. 
The integrals over $\theta$ give rise to beta-functions that can be simplified as fractions of factorials,
since the exponents of $\cos \theta$ and $\sin \theta$ are odd (this follows from the constraints). 

Before we give the final result for the transformed partition sum $Z$, we introduce linear combinations of the $j_{x,\nu}^{ab}$ as
new variables in order to make the worldline representation more transparent. The new ''flux variables'' 
$k_{x,\nu}^\lambda \in \mathds{Z}, \lambda = 1,2$ and the ''auxiliary variables'' 
$m_{x,\nu}^\lambda \in \mathds{N}_0, \lambda = 1,2$ are defined as
\begin{equation}
k_{x,\nu}^{1} = j_{x,\nu}^{11} - j_{x,\nu}^{22} \; , \; 
k_{x,\nu}^{2} = j_{x,\nu}^{12} - j_{x,\nu}^{21} \; , \; 
m_{x,\nu}^{1} = \frac{j_{x,\nu}^{11} + j_{x,\nu}^{22} - |k_{x,\nu}^1|}{2}  \; , \; 
m_{x,\nu}^{2} = \frac{j_{x,\nu}^{12} + j_{x,\nu}^{21} - |k_{x,\nu}^2|}{2} \; .
\end{equation}  
We can invert these relations and express the $j_{x\nu}^{ab}$ in terms of the $k_{x,\nu}^\lambda$ and $m_{x,\nu}^\lambda$. Inserting these 
into (\ref{trafo2}) after all integrals were solved gives rise to the final form of the worldline representation of the partition sum (compare 
\cite{kloiber} for a similar sequence of transformations),
\begin{equation}
Z \; = \; \sum_{\{k,m\}} W_J[k,m] \, W_H[k,m] \, W_\mu[k] \; 
\prod_x \prod_{\lambda = 1}^2 \, \delta \left( \vec{\nabla} \vec{k}^{\, \lambda}_x \right) \; .
\label{worldlineZ}
\end{equation}
The sum $\sum_{\{k,m\}}$ now runs over all possible configurations of the dual variables 
$k_{x,\nu}^\lambda \in \mathds{Z}, \lambda = 1,2$ and $m_{x,\nu}^\lambda \in \mathds{N}_0, \lambda = 1,2$ assigned to the links of
the lattice. While the auxiliary variables $m_{x,\nu}^\lambda$ are unconstrained, the $k_{x,\nu}^\lambda$ are subject to constraints, which 
in (\ref{worldlineZ}) are implemented by the product of Kronecker deltas (we use the notation $\delta(n) \equiv \delta_{n,0}$). These
Kronecker deltas enforce a vanishing discretized divergence for both $k_{x,\nu}^1$ and $k_{x,\nu}^2$ at every site $x$. The discretized
divergence is defined as 
$\vec{\nabla} \vec{k}^\lambda_x \equiv \sum_\nu [ k_{x,\nu}^\lambda - k_{x - \hat{\nu},\nu}^\lambda ]$ and the condition 
$\vec{\nabla} \vec{k}^\lambda_x = 0 \; \forall x$ implies that at each site $x$ the total flux of $k_{x,\nu}^\lambda$ has to vanish. In other 
words, the fluxes of $k_{x,\nu}^1$ and of $k_{x,\nu}^2$ must form closed worldlines. 

The configurations of the $k_{x,\nu}^\lambda$ and  $m_{x,\nu}^\lambda$ come with weight factors which we organize as follows:
the $J$-dependent combinatiorial weight factor $W_J[k,m]$, the weight factor $W_H[k,m]$ from the Haar measure integration 
(the above mentioned beta-functions written as fractions of factorials), and the $\mu_\lambda$-dependent weight factor $W_\mu[k]$
for the coupling to the chemical potential. The first two weight factors are 
\begin{equation}
W_J[k,m] \, = \, \prod_{x,\nu} \prod_{\lambda = 1}^2 
\frac{ J^{\, D_{x,\nu}^\lambda}}{( D_{x,\nu}^\lambda - m_{x,\nu}^\lambda)! \, m_{x,\nu}^\lambda !}
 \; , \;
W_H[k,m] \, = \, \prod_{x} \frac{ \prod_{\lambda = 1}^2 \left( \frac{1}{2}
\sum_\nu [D_{x,\nu}^\lambda + D_{x-\hat{\nu},\nu}^\lambda]\right)!}
{\left( 1+ \frac{1}{2}\sum_\nu \sum_\lambda [D_{x,\nu}^\lambda + D_{x-\hat{\nu},\nu}^\lambda]\right)!} \; ,
\label{WJWH}
\end{equation}
where we introduced the abbreviation 
\begin{equation}
D_{x,\nu}^\lambda \; \equiv \; |k_{x,\nu}^\lambda| + 2 m_{x,\nu}^\lambda \; .
\label{Dabbrev}
\end{equation}
The weight factor $W_\mu[k]$ for the $\mu$-dependence reads
\begin{equation}
W_\mu[k] \; = \; \prod_{\lambda = 1}^2 \prod_x e^{\, \mu_\lambda \, k_{x,d}^\lambda} 
\; = \; \prod_{\lambda = 1}^2e^{\, \mu_\lambda \sum_x k_{x,d}^\lambda} \; = \; 
e^{ \, \mu_1 \beta \, \omega_1[k] } \, e^{ \, \mu_2 \beta \, \omega_2[k] } \; .
\label{Wmu}
\end{equation}
In the last step we used the identity $\sum_x k_{x,d}^\lambda = N_t \, \omega_\lambda [k]$ where $\omega_\lambda [k]$ is the temporal 
net winding number of the $k^\lambda$-flux and $N_t$ the extent of the lattice in time direction, i.e., the direction $\nu = d$. 
The identity holds because admissible configurations of the $k_{x,\nu}^\lambda$ are closed worldlines. 
In (\ref{Wmu}) we also used the fact that 
$N_t$ is the inverse temperature in lattice units and replaced $N_t$ by the more conventional symbol $\beta$.

Let us conclude this section with a few remarks: Obviously all weight factors are real and positive also for non-zero $\mu_\lambda$, 
such that the worldline representation (\ref{worldlineZ}) completely solves the complex action problem. The new degrees of freedom 
are the flux variables $k_{x,\nu}^\lambda ,\, \lambda \!=\! 1,2$, subject to constraints such that admissible configurations
consist of two sets of closed worldlines for $\lambda = 1$ and $\lambda = 2$, and unconstrained configurations of 
the auxiliary dual variables 
$m_{x,\nu}^\lambda, \, \lambda = 1,2$. We remark that a worldline representation is not unique and refer to \cite{forcrand} for a different
worldline representation of the SU(2) principal chiral model. 

Via the temporal net winding numbers $\omega_\lambda [k]$ the chemical potentials $\mu_\lambda$ couple only to the 
conserved $k^\lambda$-fluxes. This is a beautiful geometrical manifestation of the underlying symmetry
and in the worldline picture we can identify the particle numbers corresponding to the Noether charges (\ref{noether}) 
as the winding numbers. Thus the particle numbers are defined as integers for individual configurations in the worldline representation, 
while in the conventional lattice representation they are non-integer functionals of the fields $U_x$ corresponding to some 
lattice discretization of the expressions (\ref{noether}). This property opens the door towards a clean and straightforward implementation
of a canonical simulations in the worldline representation which might be a more powerful approach  
in some parameter ranges (compare \cite{oliver}). 

The constraints for the $k^\lambda$-fluxes are the constraints of two U(1) subgroups since they come from integrating over 
the U(1) phases $\alpha_x$ and $\beta_x$ (the worldline form of the XY-model would have only one conserved $k$-flux). The full 
SU(2) symmetry of the model is implemented by the weight factor $W_H[k,m]$ from the Haar measure integration. It ties together the 
flux from the two U(1) worldlines at the sites of the lattice and together with the auxiliary variables $m_{x,\nu}^\lambda$ generates
the correct representation of SU(2) in the worldline form (compare also the worldline form of O(N) and CP(N-1) models discussed in
\cite{ONCPN}). Understanding the corresponding mechanism for other non-abelian 
symmetry groups or comparing it to other types of worldline representations, e.g., \cite{forcrand}, is an interesting open problem.

We remark at this point that possible worldline representations can also be constructed via the equivalence of the SU(2) principal chiral
model to the O(4) nonlinear sigma model. In this form dual representations were presented in \cite{O4A,ONCPN}. They use a different
set of dual variables with less flux degrees of freedom than the form given in (\ref{worldlineZ}). The fact that worldline 
representations are not unique and that different forms highlight different physical aspects, e.g., by making explicit the conservation of 
different charges as winding numbers, is well known. We furthermore stress that the equivalence to the O(4) model also allows one to use
the Wolff Cluster algorithm \cite{Wolffcluster} for simulations \cite{O4simul1,O4simul2}, 
although, due to the unresolved complex action issue this approach is limited to the case of vanishing chemical potential. 

\section{Kramers-Wannier dualization}

Having completed the discussion of the worldline representation we now come to the full Kramers-Wannier dualization, which we here 
present for $d = 4$ dimensions, but it is straighforward to generalize the KW-dual formulation to other dimensions. The first step is
to generate all admissible configurations of $k^\lambda$-fluxes by fluxes around plaquettes and disorder loops: We introduce plaquette 
variables $n_{x,\mu \nu}^{\, \lambda} \in \mathds{Z}$ that generate $|n_{x,\mu \nu}^{\, \lambda}|$ units of $k^\lambda$-flux around 
the plaquette $(x,\mu \nu), \, \mu < \nu$. For $n_{x,\mu \nu}^{\, \lambda} > 0$ the flux is oriented in the mathematically positive sense, 
while it has negative orientation for $n_{x,\mu \nu}^{\, \lambda} < 0$. In addition we introduce disorder loops of flux which we place on the  
coordinate axes through the origin of our $d$-dimensional lattice. Note that due to the periodic boundary conditions this gives rise to 
a closed loop of disorder flux around the periodic spatial ($\nu = 1,2 \, ... \, d-1$) and time ($\nu = d$) coordinate axes. 
The disorder flux is written as $\Theta^{(\nu)}_{x,\mu} \, q_\nu^\lambda$ where the support function 
$\Theta^{(\nu)}_{x,\mu}$ is equal to 1 for all links $(x,\mu)$ on the coordinate axis in direction $\nu$ 
and 0 for all other links. The disorder flux variables $q_\nu^\lambda \in \mathds{Z}$ introduce $|q_\nu^\lambda|$ units of $k^\lambda$-flux
on the $\nu$ coordinate axis which is oriented in positive $\nu$-direction for  $q_\nu^\lambda > 0$ and has negative orientation for 
$q_\nu^\lambda < 0$.  Thus the total flux $k^\lambda_{x,\nu}$ on the link $(x,\nu)$ is composed 
from the contributions of the plaquettes containing the link $(x,\nu)$ and the contribution from the disorder lines
if $(x,\nu)$ is on the $\nu$ coordinate axes:  
\begin{equation}
k^\lambda_{x,\nu} \; = \; 
\sum_{\rho: \nu < \rho} \Big[ n_{x,\nu \rho}^{\, \lambda} - n_{x-\hat{\rho},\nu \rho}^{\, \lambda} \Big] \; - \; 
\sum_{\mu: \mu < \nu} \Big[ n_{x,\mu \nu}^{\, \lambda} - n_{x-\hat{\mu},\mu \nu}^{\, \lambda} \Big]  \; + \; 
\Theta^{(\nu)}_{x,\nu} \, q_\nu^\lambda \; .
\label{hodge}
\end{equation}
This decomposition generates all possible configurations of $k^\lambda_{x,\nu}$ that obey the zero-divergence constraints 
(Hodge decomposition, see, e.g., \cite{wallace}). It is straightforward to rewrite the worldline formulation to the new variables:
One replaces the $k^\lambda_{x,\nu}$ in (\ref{Dabbrev}) by the representation (\ref{hodge}) and in the partition 
function (\ref{worldlineZ}) now sums over configurations of the $n_{x,\mu \nu}^{\, \lambda} \in \mathds{Z}$ and 
$q_\nu^\lambda \in \mathds{Z}$, as well as the auxiliary variables $m_{x,\nu}^{\, \lambda} \in \mathds{N}_0$. The weight factor
for the dependence on the chemical potential assumes the simple form 
$W_\mu[q] = e^{\, \mu_1 \beta \, q_d^{\, 1}} \, e^{\, \mu_2 \beta \, q_d^{\, 2}}$, since the temporal winding of $k^{\lambda}$-flux 
equals the temporal disorder variable $q_d^{\, \lambda}$.

In order to better exhibit the geometrical structure of the fully dualized theory we finally switch to the dual lattice. For
our 4-dimensional hypercubic lattice also the dual lattice is hypercubic with the same lattice spacing set to $a = 1$.  
The sites of the dual lattice are denoted by $\tilde{x}$ and are placed at the centers of the hypercubes of the original lattice. The
variables of the worldline formulation are transformed to the dual lattice as follows:
\begin{equation}
n_{x,\mu \nu}^{\, \lambda} \, \rightarrow \! \sum_{\sigma < \tau} \epsilon_{\mu \nu \sigma \tau} \, 
\widetilde{n}^{\, \lambda}_{\tilde{x} + \hat{\mu} + \hat{\nu}, \sigma \tau} \;\; (\mu \! <\!  \nu) \; , \quad
m_{x,\nu}^{\, \lambda} \, \rightarrow \!\! \sum_{\sigma < \tau < \omega} \!\! \epsilon_{\nu \sigma \tau \omega} \, 
\widetilde{m}^{\, \lambda}_{\tilde{x} + \hat{\nu} , \sigma \tau \omega} 
\; , \quad
\Theta^{\; (\nu)}_{x,\nu} \, q_\nu^{\; \lambda} \, \rightarrow \, 
\widetilde{\Theta}^{\;(\tilde{\nu})}_{\, \tilde{x} + \hat{\nu},\tilde{\nu}} \; \widetilde{q}_{\; \tilde{\nu}}^{\; \, \lambda} \; .
\end{equation}
Here $\epsilon_{\mu \nu \sigma \tau}$ is the completely antisymmetric tensor and we use the notation $\tilde{\nu}$ to label
the cubes of the dual lattice which are dual to links in direction $\nu$ of the original lattice. In other words $\tilde{\nu}$ is the triplet 
of three coordinates obtained by deleting $\nu$ in the 4-tuplet (1,2,3,4). 
Consequently $\widetilde{\Theta}^{\;(\tilde{\nu})}_{\, \tilde{x},\tilde{\mu}}$ is the support function on the dual lattice which is 1 for all
dual cubes $(\tilde{x}, \tilde{\mu})$ that are dual $\tilde{\nu}$-cubes on the $\nu$-axis (i.e., the cubes dual to the $\nu$-axis of the original 
lattice) and 0 for all other cubes of the dual lattice. 

In the KW-dual formulation we thus use the dual dynamical variables $\widetilde{n}^{\, \lambda}_{\tilde{x}, \sigma \tau} \in \mathds{Z}$ 
assigned to the plaquettes of the dual lattice, the dual auxiliary variables 
$\widetilde{m}^{\, \lambda}_{\tilde{x}, \tilde{\nu}} \in \mathds{N}_0$ on the cubes of the dual lattice, and the dual disorder variables
$\widetilde{q}_{\; \tilde{\nu}}^{\; \, \lambda} \in \mathds{Z}$ for the flux through the dual cubes dual to the coordinate axes. 
After some algebra one obtains the fully KW-dual form of the partition sum:
\begin{equation}
Z \; = \; \sum_{\tilde{q}^{\,1}_{\tilde{4}}, \, \tilde{q}^{\, 2}_{\tilde{4}} \in \mathds{Z}} 
e^{\, \beta \left(\mu_1 \, \tilde{q}_{\tilde{4}}^{\, 1} + \mu_2 \, \tilde{q}_{\tilde{4}}^{\, 2} \right)}
\left( \prod_{\nu = 1}^3 \; \sum_{\tilde{q}^{\,1}_{\tilde{\nu}}, \, \tilde{q}^{\,2}_{\tilde{\nu}} \in \mathds{Z} }\right)
\sum_{\{ \tilde{n}, \tilde{m} \}} W_J[\widetilde{n}, \widetilde{q}, \widetilde{m}] \, 
W_H[\widetilde{n}, \widetilde{q}, \widetilde{m}] \; ,
\label{KWdual}
\end{equation}
where the sum $\sum_{\{ \tilde{n}, \tilde{m} \}}$ now runs over all configurations of the dual 
$\widetilde{n}$- and $\widetilde{m}$-variables on the 
dual lattice. In (\ref{KWdual}) the sums over the dual disorder variables 
$\widetilde{q}^{\, \lambda}_{\tilde{\nu}}$ were written explicitly up front. They are
ordered such that the first double sum is over the temporal disorder variables $\widetilde{q}^{\, \lambda}_{\tilde{4}}$ 
which carry the dependence on the chemical potentials $\mu_\lambda$. Thus the KW-dual partition sum 
(\ref{KWdual}) is already organized in the form of a double fugacity expansion. 
The weight factor $W_J[\widetilde{n}, \widetilde{q}, \widetilde{m}]$ from the Taylor expansion 
of the original Boltzman factors reads
\begin{equation}
W_J[\widetilde{n}, \widetilde{q}, \widetilde{m}] \, = \, \prod_{\tilde{x},\tilde{\nu}} \prod_{\lambda = 1}^2 
\frac{ J^{\, \widetilde{D}_{\tilde{x},\tilde{\nu}}^\lambda}}{( \widetilde{D}_{\tilde{x},\tilde{\nu}}^\lambda - 
\widetilde{m}_{\tilde{x},\tilde{\nu}}^\lambda)! \, \widetilde{m}_{\tilde{x},\tilde{\nu}}^\lambda !} \; ,
\end{equation}
where the combinations $\widetilde{D}_{\tilde{x},\tilde{\nu}}^\lambda \in \mathds{N}_0$ assigned to the dual cubes $(\tilde{x},\tilde{\nu})$ 
are given by 
\begin{equation}
\widetilde{D}_{\tilde{x},\tilde{\nu}}^\lambda \, =  
\left| \, \sum_{\sigma < \tau} \! \left( 
\sum_{\rho: \nu < \rho} \!\!\! \epsilon_{\nu \rho \sigma \tau} 
\Big[ \widetilde{n}^{\, \lambda}_{\tilde{x} + \hat{\rho}, \sigma \tau} - \widetilde{n}^{\, \lambda}_{\tilde{x}, \sigma \tau} \Big] - \! \!
\sum_{\mu: \mu < \nu} \!\!\! \epsilon_{\nu \mu \sigma \tau} 
\Big[ \widetilde{n}^{\, \lambda}_{\tilde{x} + \hat{\mu}, \sigma \tau} - \widetilde{n}^{\, \lambda}_{\tilde{x}, \sigma \tau} \Big] \! \right) \!
+ \widetilde{\Theta}^{\;(\tilde{\mu})}_{\, \tilde{x},\tilde{\nu}} \; \widetilde{q}_{\; \tilde{\nu}}^{\; \, \lambda} \, \right| \, + \, 
2 \widetilde{m}_{\tilde{x},\tilde{\nu}}^\lambda \; .
\end{equation}
The weight $W_H[\widetilde{n}, \widetilde{q}, \widetilde{m}]$ that originates from the Haar measure integration and implements the 
SU(2) symmetry of the conventional representation in the KW-dual formulation is given by
\begin{equation}
W_H[\widetilde{n}, \widetilde{q}, \widetilde{m}] \; = \; \prod_{\tilde{h}} 
\frac{ \prod_{\lambda} 
\left( \frac{1}{2} \sum_{(\tilde{x}, \tilde{\nu}) \in \partial \tilde{h}} \! \widetilde{D}_{\tilde{x},\tilde{\nu}}^\lambda \right) ! }{
\left( 1 + \frac{1}{2} 
\sum_{\lambda} \sum_{(\tilde{x}, \tilde{\nu}) \in \partial \tilde{h}} \! \widetilde{D}_{\tilde{x},\tilde{\nu}}^\lambda \right) ! } \; ,
\label{WHKW}
\end{equation}
where the product $\prod_{\tilde{h}}$ runs over all hypercubes $\tilde{h}$ of the dual lattice, and the sum 
$\sum_{(\tilde{x}, \tilde{\nu}) \in \partial \tilde{h}}$ is over all dual cubes $(\tilde{x}, \tilde{\nu})$ in the boundary $\partial \tilde{h}$
of $\tilde{h}$. In the KW-dual form all constraints have disappeared and again all weights are real and positive, such that a 
Monte Carlo simulation is possible at finite $\mu^\lambda$. 

We conclude the discussion of the KW-dual form by pointing out that it is dual also in the sense that the weak and strong coupling 
limits are interchanged: In the conventional form the ordered phase with constant $U_x$ appears at large couplings, 
while in the KW-dual form small $J$ favors an ordered phase with constant values of the dual variables 
$\widetilde{n}^{\, \lambda}_{\tilde{x}, \sigma \tau}$. This aspect could be explored when using the different formulations
for numerical simulations in different coupling regimes of the model.

\section{An exploratory numerical test}

The worldline form (\ref{worldlineZ}) -- (\ref{Wmu}) and the KW-dual form (\ref{KWdual}) -- (\ref{WHKW}) are the two formulations 
which we were aiming at in this letter. We now present results of first exploratory numerical tests of the two formulations
in $d = 4$ dimensions. They mainly serve to test the duality transformations, but also shed a first light on the potential for  
using the worldline- and the KW-form for Monte Carlo simulations. 

More specifically we implemented four Monte Carlo simulations for cross-checking the derivation of the worldline- and 
KW-dual formulations: The first simulation is a simulation at $\mu_\lambda = 0$ directly in the conventional representation
(\ref{latticeaction}), which is free of the complex action problem at vanishing chemical potentials. This simulation serves as
a reference case for numerically testing the worldline- and KW-dual formulations on $16^4$ lattices (heat bath updates, 
$10^4$ decorrelated measurements per coupling). In the worldline formulation we performed two types of simulations: One is
an update with a generalization of the worm algorithm \cite{worm} (taking into account the site terms using the variant described in
\cite{mario}), the other one a local update corresponding to the decomposition (\ref{hodge}): one sweep consists of offering a 
change of the flux by one unit for all plaquettes, followed by offering the change of flux along straight loops that close around the
periodic boundary conditions. For this local update in the worldline form we used $2.5 \times 10^4$ configurations decorrelated 
by 5 combined sweeps for the $\mu = 0$ runs and up to $10^5$ configurations for the runs at non-zero chemical potential. The
worm algorithms turned out to be efficient only below the transition in $J$ (see the discussion below) and we used statistics of 
$2.5 \times 10^4$ worms in that region. Finally, in the fully KW-dual formulation we implemented sweeps of local Metropolis steps, again 
using between $10^4$ and $10^5$ configurations. The auxiliary variables were always updated with local Metropolis sweeps, both in 
the worldline- as well as the KW-dual formulation.

\begin{figure}[t]
\begin{center}
\hspace*{1mm}
\includegraphics[height=5.2cm]{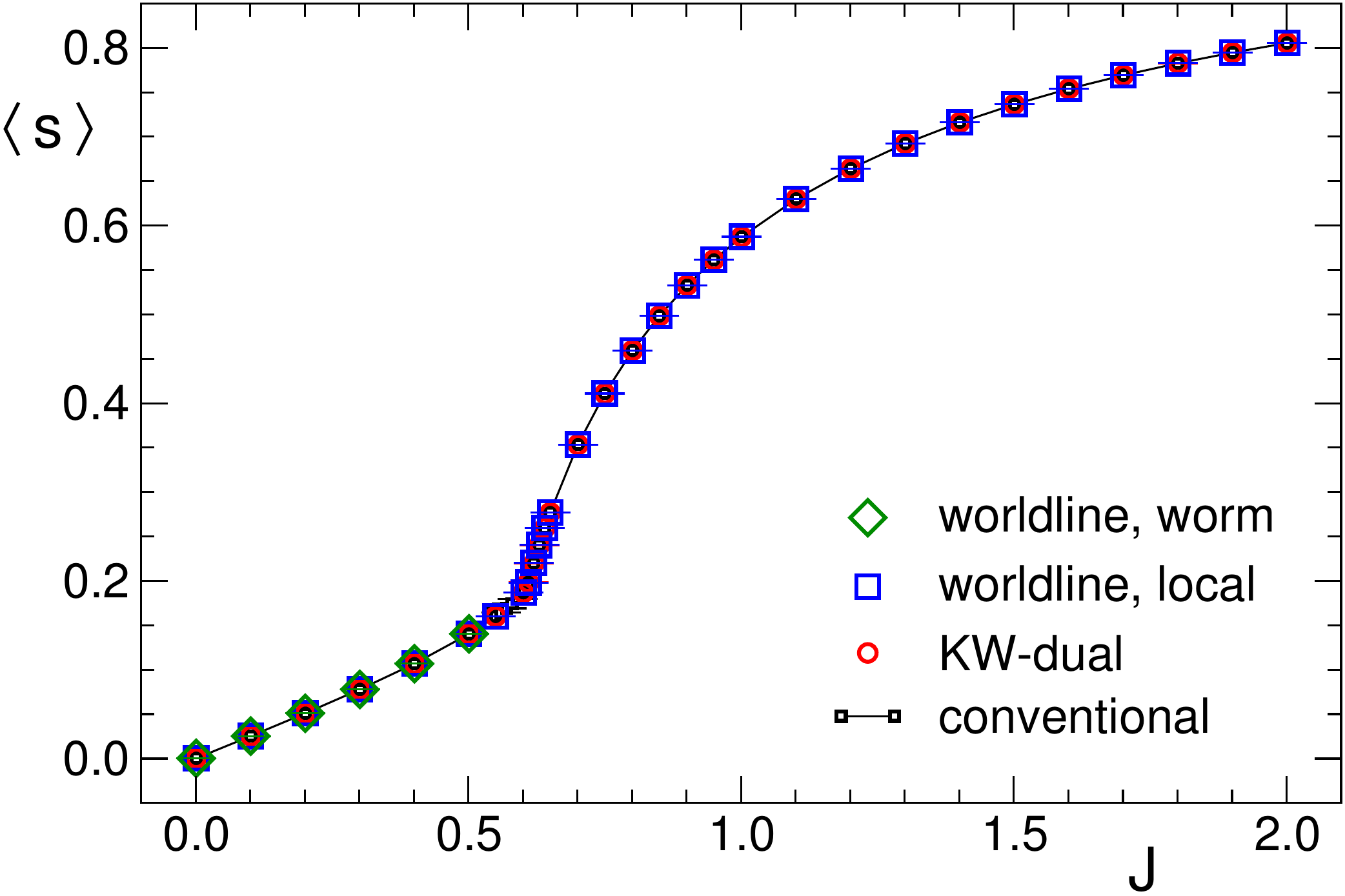}
\hspace{2mm}
\includegraphics[height=5.2cm]{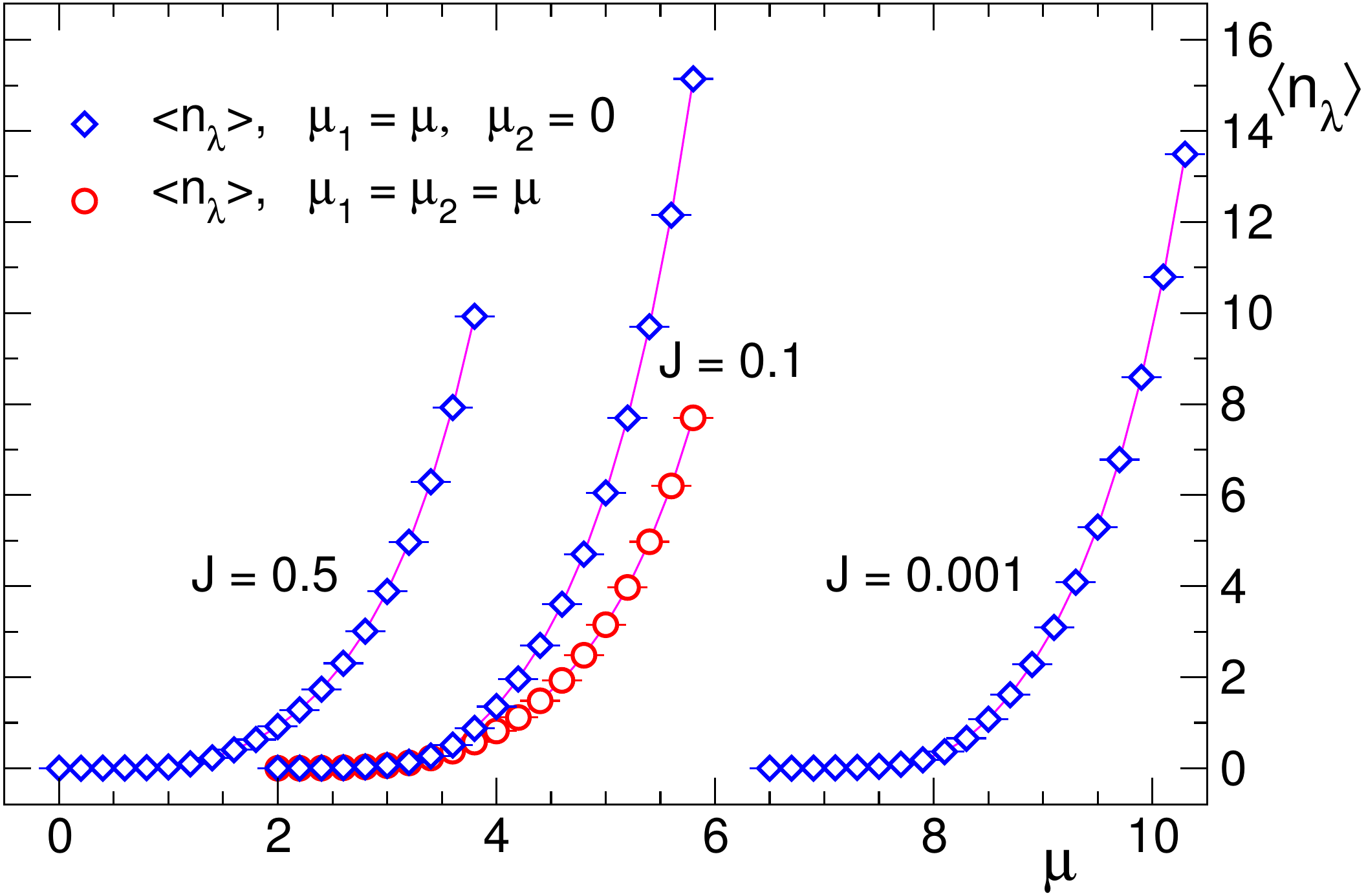}

\includegraphics[height=3.77cm]{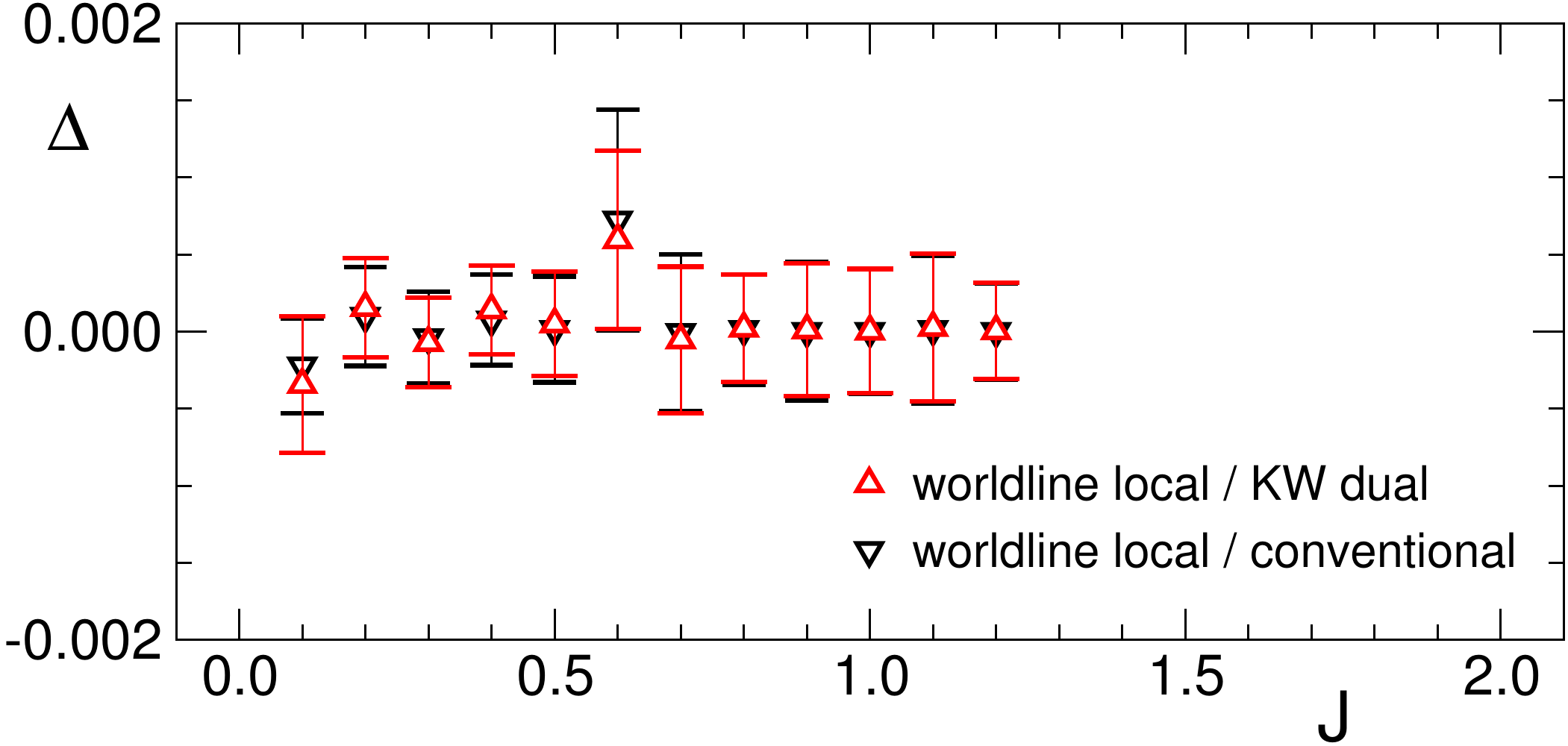}
\hspace{1.98mm}
\includegraphics[height=3.67cm]{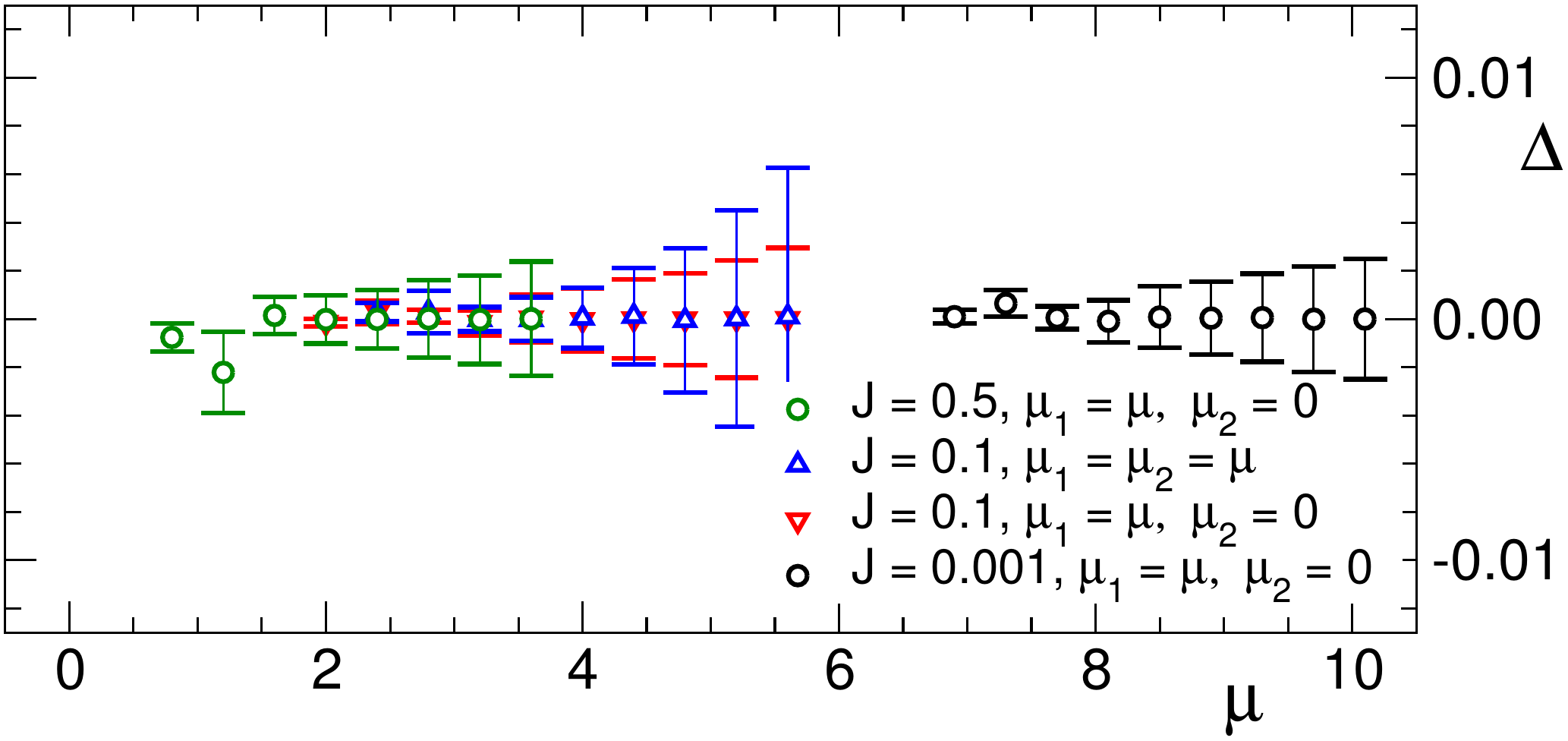}
\end{center}
\vspace*{-6mm}
\caption{Lhs.~top plot: The action density $\langle s \rangle$ as a function of the coupling $J$ ($\mu_1 = \mu_2 = 0$, lattice size $16^4$). 
We compare the results from the simulation in the conventional representation (small squares), two simulations of the worldline 
representation (results from local updates are represented as large squares, worm simulation results are shown as diamonds), 
as well as the results from a simulation of the KW-dual form (circles). 
Rhs.\ top plot: Particle number densities $\langle n_\lambda \rangle$ as a function of the chemical potential $\mu$. Using a $16^3 \times 4$ 
finite temperature lattice the couplings $J = 0.001$, 0.1 and $J = 0.5$ were studied. We analyze two 
different scenarios:  $\langle n_\lambda 
\rangle$ versus $\mu_1 = \mu$ at 
$\mu_2 = 0$ (diamonds), as well as $\langle n_\lambda \rangle$ versus $\mu_1 = \mu_2 = \mu$ (circles, only for $J$ = 0.1). 
We compare the results of the local worldline representation (no symbols, data connected with lines) to the KW-dual results 
(symbols). Bottom plots: The relative errors $\Delta = (y_1 - y_2)/y_1$ for a comparison of the results 
$y_1, y_2$ determined with different methods. The observables are the same as shown in the plots above.}
\label{simulationplots}
\end{figure}

The observables we considered were bulk quantities obtained as derivatives of $\ln Z$ with respect to the parameters of the theory.
More specifically we studied the action density $\langle s \rangle$ and the particle number densities $\langle n_\lambda \rangle$ 
defined as
\begin{equation}
\langle s \rangle \; = \; \frac{1}{8 N^3 N_t} \, \frac{\partial}{\partial J} \, \ln Z   \quad , \quad 
\langle n_\lambda \rangle \; = \; \frac{1}{N^3 N_t} \, \frac{\partial}{\partial \mu_\lambda} \, \ln Z  \; .
\end{equation}
These derivatives can be applied to the partition function $Z$ in any of the three representations, and the observables
are obtained as sums over local expressions of the degrees of freedom. For the example of the action density this is the expectation 
value of the sum over traced nearest neighbor terms $U_x U_{x+\hat{\nu}}^\dagger$, the sum over the combinations $D_{x,\nu}^\lambda$
for the worldline formulation, and the sum over the $\widetilde{D}_{\tilde{x},\tilde{\nu}}^{\, \lambda}$ in the KW-dual form. As already 
remarked, for the particle number densities the worldline representation is the temporal winding number of the worldlines and in the 
KW-dual form the corresponding dual sum. 

In the top row of Fig.~\ref{simulationplots} we show the results of our exploratory numerical simulations. In the lhs.\ top
plot we display the action density as
a function of the coupling $J$ at $\mu_1 = \mu_2 = 0$ on a $16^4$ lattice. We compare the results from the simulation in 
the conventional formulation (small squares), the worm update (diamonds) and a local simulation (large squares) in the 
worldline formulation, as well as the simulation of the KW-dual form (circles). The curves for the four data sets fall perfectly on top of 
each other and demonstrate that the worldline and KW-dual forms were derived correctly and are suitable for numerical simulations.  
As we discuss below, the worm becomes rather inefficient in the ordered phase ($J > 0.62$) and the corresponding data thus end at 
$J = 0.5$. 

In the rhs.\ top plot we show the particle number densities $\langle n_\lambda \rangle$ as a function of the chemical potential parameter 
$\mu$ using a finite temperature lattice of size $16^3 \times 4$ and three different values of the coupling: $J = 0.001, 0.1$ and
$J = 0.5$. Here we compare the results from the local worldline simulation 
(no symbols, data connected with lines) to the data from the KW-dual form (different symbols - see labelling of the data sets). 
We considered different scenarios: $\mu_1 = \mu_2 = \mu$ and $\mu_1 = \mu, \, \mu_2 = 0$ and show 
$\langle n_\lambda \rangle$ versus $\mu$. In both cases we see the onset of condensation for both 
densities $\langle n_\lambda \rangle$ with the condensation threshold for $\mu$ depending on the coupling $J$. In both scenarios 
and for all values of the coupling $J$ we again find perfect agreement of the data, thus successfully testing the dual representations
also at non-zero $\mu_\lambda$. 

For a more detailed comparison of the results from different methods, in the bottom row of plots of
Fig.~\ref{simulationplots} we show the relative errors $\Delta = (y_1 - y_2)/y_1$  for pairs $y_1, y_2$ of 
results determined with different methods. For this comparison the statistics was increased by a factor of 10.
The statistical errors we show were determined with the jackknife method combined with a blocking analysis.  
In the lhs.\ bottom plot we show the relative error for $\langle s \rangle$ and 
compare the KW dual and the conventional results with the reference data from the local worldline simulation. The  worldline simulation with worms was left out in this comparison due to bad performance (see above). 
The relative errors are small and we find very good agreement of the results from KW dual, local worldline and
conventional simulations. Similar good agreement is seen in the 
rhs.\ plot for the relative error of $\langle n_\lambda \rangle$, comparing KW-dual and local worldline updates. 
In summary the agreement of the
different methods illustrated in the bottom plots of Fig.~\ref{simulationplots} is very convincing.

We have already remarked that the Monte Carlo simulations presented here mainly serve to check the worldline and KW-dual 
representations and it is clear that for a proper assessment of the power of the different formulations a much more detailed numerical 
analysis is necessary. Nevertheless we would like to comment on our finding that in this first exploratory study the worm did not perform
very well in the ordered phase (the region $J > 0.62$ in the lhs.\ plot of Fig.~\ref{simulationplots}) and at finite $\mu_\lambda$. As discussed
above, for the worldline formulation of the principal chiral model we have also weights at the sites of the lattice such that the worm 
described in \cite{mario} has to be used, where the site weights are treated correctly. This worm also contains an amplitude parameter that 
can be chosen freely and allows one to adapt the starting and terminating probabilities of the worm. Despite carefully fine-tuning this 
parameter we could not find a window of decent performance of the worm for some coupling regions. Either worms turned 
out to be very short, often simply retracing their initial step, and thus being less efficient than the local or KW-dual updates, 
or, in particular for $\mu_\lambda > 0$, very long and hardly ever closing worms resulted, again leading to an inefficient algorithm. 
Certainly these findings will have to be tested more carefully, but one can already conclude that for some coupling regions
the local worldline updates or simulating the KW-dual form is more efficient.  

\section{Summary and discussion}

In this letter we have presented a worldline formulation and a fully Kramers-Wannier dualized form of the SU(2) principal chiral model 
with chemical potentials coupled to two of the conserved charges. The derivation of the worldline formulation is based on the Abelian 
color flux approach where the matrix products and traces over the group indices are written explicitly and corresponding expansion
variables are introduced for each term, which then will turn into the dual variables. Integrating out the original degrees of 
freedom gives rise to constraints for the new variables, which here transform the partition function into a sum over two sets of 
worldlines with interaction terms on the sites that implement the SU(2) symmetry of the original formulation together with auxiliary dual 
variables free of constraints. The chemical potentials couple to the temporal winding numbers of the two sets of worldlines, thus
giving the particle number an elegant geometrical form and solving the complex action problem of the original formulation. 
The KW-dual form can subsequently be derived by decomposing the worldlines 
into local plaquette fluxes and disorder loops that wind around the lattice. We stress at this point that the strategy of using the abelian 
color flux approach to first obtain a worldline- and then the fully KW-dual form is rather general and we currently 
explore its application to other 
non-abelian symmetry groups. A small numerical simulation for testing the worldline- and KW dual formulations completes the paper. 

There are several aspects concerning the motivation and the perspective of the study presented here: For non-abelian theories relatively
little is known about fully dual representations, and the approach presented here might be useful for further development in this direction. 
Coupling the chemical potentials $\mu_\lambda$ is not only interesting because the worldline- and KW-dual forms completely solve 
the complex action problem, but also sheds light on another important aspect: The chemical potentials are coupled to the conserved 
charges that correspond to symmetries of the model in the conventional simulation. Thus analyzing how the chemical potential appears
in the new representations allows one to monitor how the original symmetries are manifest in the worldline- or KW-dual formulations. 

Finally, also for actual numerical finite density simulations of systems, which in the conventional approach have a complex action problem, 
the systematical study of different representations is important. For many models worldline representations were used in recent years
to study their physics at non-zero chemical potential. Although the complex action problem is solved with finding a real and positive
worldline form, the numerical simulation still can suffer from severe autocorrelation problems. The preliminary numerical findings 
presented here indicate that in some coupling regions the worm update is inefficient despite fine tuning the worm amplitude parameter. 
In these cases switching to a completely KW-dual form is a good choice for efficient simulations, and the absence of 
constraints in the KW-dual form might even allow for using strategies such as Swendsen-Wang type algorithms. These are 
questions that will be addressed in more detail in future work.

\vskip5mm
\noindent
{\bf Acknowledgements:} We thank Falk Bruckmann, Tin Sulejmanpasic and Ulli Wolff for interesting discussions. 
Daniel G\"oschl and Carlotta Marchis are partly funded by the FWF DK W1203, "Hadrons in Vacuum, Nuclei and Stars". 
Furthermore this work is supported by the Austrian Science Fund FWF, grant number I 2886-N27, as well as 
DFG TR55, ''$\!$Hadron Properties from Lattice QCD''.

\end{document}